\documentclass[fleqn,usenatbib]{mnras}

\usepackage{newtxtext,newtxmath}

\usepackage[T1]{fontenc}
\usepackage{ae,aecompl}

\usepackage{graphicx}	
\usepackage{amsmath}	
\usepackage{amssymb}	
\usepackage[usenames,dvipsnames]{xcolor} 
\usepackage[normalem]{ulem} 
\usepackage{hyperref}

\usepackage{lipsum}
\usepackage{multirow}

\newcommand{\ceagle}{\mbox{\sc{C-Eagle}}}
\newcommand{\eagle}{\mbox{\sc{Eagle}}}
\newcommand{\spitzer}{\mbox{\it Spitzer}}

\newcommand{\hubble}{\mbox{\it Hubble}}

\newcommand{\webb}{\mbox{\it JWST}}

\newcommand{\flares}{\mbox{\sc Flares}}

\definecolor{deepchestnut}{rgb}{0.73, 0.31, 0.28}

\title[FLARES VII: Star Formation and Metal Enrichment]{First Light And Reionisation Epoch Simulations (FLARES) VII: The Star Formation and Metal Enrichment Histories of Galaxies in the early Universe}

\author[Stephen M. Wilkins et al.]{Stephen M. Wilkins$^{1,2}$\thanks{E-mail: s.wilkins@sussex.ac.uk}, 
Aswin P. Vijayan$^{3,4,1}$, 
Christopher C. Lovell$^{5,1}$, 
William J. Roper$^{1}$,\newauthor  
Erik Zackrisson$^{6}$, 
Dimitrios Irodotou$^{7,1}$, 
Louise T. C. Seeyave$^{1}$, 
Jussi K. Kuusisto$^{1}$, 
Peter A. Thomas$^{1}$, 
\newauthor
Joseph Caruana$^{2,8}$,  
Christopher J. Conselice$^{9}$
\\
$^{1}$Astronomy Centre, University of Sussex, Falmer, Brighton BN1 9QH, UK\\
$^{2}$Institute of Space Sciences and Astronomy, University of Malta, Msida MSD 2080, Malta \\
$^{3}$Cosmic Dawn Center (DAWN) \\
$^{4}$DTU-Space, Technical University of Denmark, Elektrovej 327, DK-2800 Kgs. Lyngby, Denmark \\
$^{5}$Centre for Astrophysics Research,  School of Physics, Engineering \& Computer Science, University of Hertfordshire, Hatfield AL10 9AB, UK\\
$^{6}$ Observational Astrophysics, Department of Physics and Astronomy, Uppsala University, Box 516, SE-751 20 Uppsala, Sweden\\
$^{7}$Department of Physics, University of Helsinki, Gustaf Hällströmin katu 2, FI-00014, Helsinki, Finland\\
$^{8}$Department of Physics, University of Malta, Msida MSD 2080, Malta \\
$^{9}$Jodrell Bank Centre for Astrophysics, University of Manchester, Oxford Road, Manchester, UK\\
}

\date{Accepted XXX. Received YYY; in original form ZZZ}

\pubyear{2022}

\begin{document}
\label{firstpage}
\pagerange{\pageref{firstpage}--\pageref{lastpage}}
\maketitle

\begin{abstract}
The star formation and metal enrichment histories of galaxies - at any epoch - constitute one of the key properties of galaxies, and their measurement is a core aim of observational extragalactic astronomy. The lack of deep rest-frame optical coverage at high-redshift has made robust constraints elusive, but this is now changing thanks to the \emph{James Webb Space Telescope (JWST)}. In preparation for the constraints provided by \emph{JWST} we explore the star formation and metal enrichment histories of galaxies at $z=5-13$ using the First Light And Reionisation Epoch Simulations (\flares) suite. Built on the \eagle\ model, the unique strategy of \flares\ allows us to simulate a wide range of stellar masses (and luminosities) and environments. While we predict significant redshift evolution of average ages and specific star formation rates our core result is a mostly flat relationship of age and specific star formation rate with stellar mass. We also find that galaxies in this epoch predominantly have strongly rising star formation histories, albeit with the magnitude dropping with redshift and stellar mass. In terms of chemical enrichment we predict a strong stellar mass - metallicity relation present at $z=10$ and beyond alongside significant $\alpha$-enhancement. Finally, we find no environmental dependence of the relationship between age, specific star formation rate, or metallicity with stellar mass.

\end{abstract}

\begin{keywords}
methods: numerical -- galaxies: formation -- galaxies: evolution -- galaxies: high-redshift -- galaxies: extinction -- infrared: galaxies
\end{keywords}



\section{Introduction}\label{sec:intro}

A key goal of observational extragalactic astrophysics is the measurement of star formation and metal enrichment histories of representative samples of galaxies stretching over a wide range of redshifts and stellar masses. Doing so provides critical insights into the physical processes responsible for galaxy formation and evolution \citep[e.g][]{schaye_physics_2010}. These histories can also be used to measure the cosmic history star formation and metal enrichment history, and the build up of stellar mass throughout cosmic history. This includes probing star formation which is observationally inaccessible in-situ, such as in small galaxies at very-high redshift.

While much progress has been achieved at low and intermediate redshift, particularly thanks to surveys like SDSS, GAMA, COSMOS, UltraVISTA, VIDEO, and CANDELS \citep[e.g.][]{Adams21, McLeod2021, Driver22}, constraints at the highest redshifts ($z>5$) remain highly uncertain for several reasons. Firstly, most high-redshift galaxies, at least those identified so far, have intense recent star formation. Young stellar populations in these systems dominate the energy output, “outshining” older stellar populations, even in the rest-frame optical. Secondly, prior to \webb, our only access to the rest-frame optical at sensitivities unattainable by ground based telescopes, was from \spitzer, and then only typically in two broad photometric bands at 3.6$\mu$m and 4.5$\mu$m. The use of \spitzer\ not only limits us to the brightest sources but also limits \emph{clean} access to essential diagnostics, including the Balmer break and the strong optical emission lines. This is further complicated by the increasing prevalence of strong nebular line emission \citep[e.g.][]{Wilkins2013, Stark2013, deBarros2014, Smit2015, Wilkins2020} at high-redshift which can be easily confused for age sensitive features like the Balmer break, especially where only photometric redshifts are available. Despite these challenges, however, there has been some progress in constraining the stellar masses and star formation histories of galaxies at $z>5$ prior to \webb, \citep[e.g][]{Eyles07, CurtisLake13, Duncan14, Salmon2015, Grazian15, Katsianis15, Song16, Davidzon17, Bhatawdekar19, Strait20, Santos20, Endsley21, Strait21, Laporte21, Stefanon22, Tacchella22, Whitler2022, Topping22}, with some constraints now available even at $z\sim 10$ \citep{Laporte21, Tacchella22}. These have, so far, revealed a consistent picture of falling stellar mass densities and increasing specific star formation rates to higher-redshift. However, finer details, such as the slope of the specific star formation rate - stellar mass relation, remain highly uncertain.

With the successful commissioning of the highly sensitive \emph{James Webb Space Telescope} we can now address many open questions beyond \spitzer's capabilities. \webb\ not only provides sensitive near and mid-infrared imaging in several bands, extending to $5\mu$m and beyond with MIRI, but also near-infrared spectroscopy. Spectroscopy will provide unambiguous redshifts, allow us to constrain the contribution of nebular line emission, and enable the use of powerful rest-frame optical emission line diagnostics. At the time of writing, the first constraints on the star formation and metal enrichment histories of high-redshift galaxies from \webb\ are emerging \citep[e.g.][]{Carnall22, Naidu2022, Leethochawalit22, Trump2022, Finkelstein2022b, Chen2022}. 

Anticipating the power of these observations, in this work we present comprehensive predictions for the star formation and metal enrichment histories of galaxies simulated by \flares: First Light And Reionisation Epoch Simulations project \citep{FLARES-I, FLARES-II}. \flares\ combines the calibrated \eagle\ physics model \citep{EAGLE, EAGLE_calib} with a unique simulation strategy designed to efficiently simulate galaxies over a wide range of masses and environments at high-redshift. This wider dynamic range is key to capturing the range of galaxy populations accessible to \webb.

This paper is structured as follows, in Section \ref{sec:flares} we introduce the \flares\ suite of simulations. In Section \ref{sec:sf} we then explore \flares\ predictions for the star formation histories of galaxies. Here we make predictions using several different metrics (\S\ref{sec:sf.metrics}), fit individual star formation histories by a range of simple parameterisations (\S\ref{sec:sf.parameterisation}), compare with existing observations (\S\ref{sec:sf.observations}), and explore the environmental dependence (\S\ref{sec:sf.environment}). In Section \ref{sec:Z} we then focus on the metal enrichment of galaxies. We first present the evolution mass-metallicity relation (\S\ref{sec:Z.MZR}), including its environmental dependence (\S\ref{sec:Z.MZR.delta}). We then explore how stellar metallicity is correlated with age (\S\ref{sec:Z.age}) and the distribution of metallicities within individual galaxies  (\S\ref{sec:Z.dist}). Finally, in Section \ref{sec:conc} we present our conclusions. In this work distance measures preceded by `c' are in comoving units while the ones with `p' are in physical units. We assume a Planck year 1 cosmology \cite[$\Omega_{\mathrm{m}}=0.307$, $\Omega_{\Lambda}=0.693$, h$=0.6777$,][]{planck_collaboration_planck_2014}.

\section{The First Light And Reionisation Epoch Simulations}\label{sec:flares}

In this study, we make use of the \flares: First Light And Reionisation Epoch Simulations simulation suite. \flares\ is introduced in \citet{FLARES-I} and \citet{FLARES-II} and we direct the reader to those papers for a detailed introduction. In brief, \flares\ is a suite of 40 spherical $14\ h^{-1}\, {\rm cMpc}$ radius re-simulations. Regions re-simulated by \flares\ are selected from a large $(3.2\ {\rm cGpc})^3$ parent dark matter only simulation and span a range of environments: (at $z\approx 4.7$) $\log_{10}(1+\delta_{14}) = [-0.3, 0.3]$\footnote{Where $\delta_{14}$ is the density contrast measured within the re-simulation volume size.} with over-representation of the extremes. With the knowledge of each region's density contrast, combined distribution functions can be constructed which approximate a much larger volume than that simulated, allowing \flares\ to predict distribution functions over a much larger dynamic range than that achievable with hydrodynamic periodic volumes.

\subsection{Physics Model}

\flares\ adopts the AGNdT9 variant of the \eagle\ simulation project \cite[][]{EAGLE, EAGLE_calib} and utilises identical resolution and cosmological parameters to the fiducial \eagle\ reference simulation. The AGNdT9 variant produces similar mass functions to the reference model but better reproduces the hot gas properties in groups and clusters leading to its utilisation by the \ceagle\ project \citep{Barnes2017}. The key physics of the \eagle\ model relevant to this work concerns star formation and chemical enrichment and here we briefly summarise these, deferring a thorough introduction to  \citet[][]{EAGLE} and \citet{EAGLE_calib} and references therein. 

\subsubsection{Star Formation}

Current large cosmological simulations, including \eagle, do not have the resolution to simulate star formation from first principles. In \eagle\ the star formation recipe uses the observed Kennicutt-Schmidt star formation law \cite[]{KSlaw1998}, now rewritten as a metallicity dependent pressure law \cite[equation 1 in][]{EAGLE}. This is implemented as described in \cite{Schaye_DallaVecchia2008}, where gas particles which are cold enough (with temperatures $\lesssim 10^4$ K) above the metallicity-dependent star formation threshold of \cite{Scahye2004} are stochastically converted to star particles using the pressure-dependent version of the observed Kennicutt-Schmidt law. The star particles formed are assumed to represent a simple stellar population (SSP) formed with a \cite{chabrier_imf_2003} initial  mass function.

\subsubsection{Chemical Enrichment}

These SSPs lose mass through stellar winds arising from asymptotic giant branch (AGB) stars and massive stars as well as type Ia (SNIa) and type II (SNII) supernovae, following the prescription of \cite{Wiersma2009b}. Stellar particles lose mass dependent on their age and metallicity due to the main sequence lifetimes of the constituent stars \cite[]{Portinari1998,Marigo2001}. This mass is distributed to neighbouring gas particles based on the SPH kernel. The elements H, He, C, N, O, Ne, Mg, Si, and Fe are tracked individually. At early times ($\lesssim 100$ Myr), the mass loss is mainly from SNII, leading to significant $\alpha$-enhancement (i.e. [$\alpha$/Fe]$>0$) in younger galaxies. Later the mass loss comes from AGB and SNIa.

Most massive galaxies ($M_\star/M_\odot > 10^9$) at $z>5$ are compact with the majority of star formation taking place in dense cores \citep{Roper22}. The low metallicity present at high redshift in tandem with the core density yield strong enrichment in these cores, while enrichment is inhibited by low densities elsewhere. Nonetheless, it should be noted that despite the enhancement, stellar feedback is still inefficient at high redshift.

\eagle\ tracks both the particle abundance (directly enriched by star particles) and the smoothed abundance \cite[smoothed value obtained using the SPH kernel, see][]{Wiersma2009b}. Since the simulation does not implement diffusion of metals between gas particles, it can lead to certain individual particles exhibiting extreme metallicities. All the quoted metallicity values in this work are smoothed metallicities to mitigate some effects of this. The smoothed abundances were also used in computing the cooling rates of the gas, while the probability of star formation used the particle metallicities.

\subsection{Galaxy Identification}

Galaxies in \flares\ are first identified as groups via the Friends-Of-Friends \citep[FOF,][]{davis_evolution_1985} algorithm, and subsequently subdivided into bound objects with the \textsc{Subfind} \citep{springel_populating_2001,dolag_substructures_2009} algorithm. For a full description of the method and handling of pathological objects see \citet{McAlpine2016}, whose methodology we follow.

In this work we restrict ourselves to galaxies with $M_{\star}>10^{8.5}\ {\rm M_{\odot}}$ corresponding to a minimum of a few hundred star particles. By default, when measuring properties we use 30 pkpc radius apertures, centred on the most bound particle of each subgroup (the particle with the highest gravitational potential). However, in \S\ref{sec:appendix.apertures} we explore the consequences of this assumption: while the choice of aperture can have a significant impact on the total stellar mass and star formation, the impacts on the specific star formation rate (and other metrics of the star formation history) mostly cancel out.

\section{Star Formation Histories}\label{sec:sf}

The star formation history (SFH) describes the evolution of a galaxy's star formation activity. In the case of hydrodynamical simulations like \flares\ the SFH is defined by the age (and initial mass) distribution of the star particles making up the galaxy. To describe the SFH of a galaxy we can employ several different metrics, including: the specific star formation rate, age, and moments, amongst others. In this section we begin by presenting stacked star formation histories before presenting predictions for these metrics, then go on to explore their environmental dependence, and compare with observations.

\subsection{Stacked star formation histories}\label{sec:sf.stacked}

\begin{figure}
	\includegraphics[width=\columnwidth]{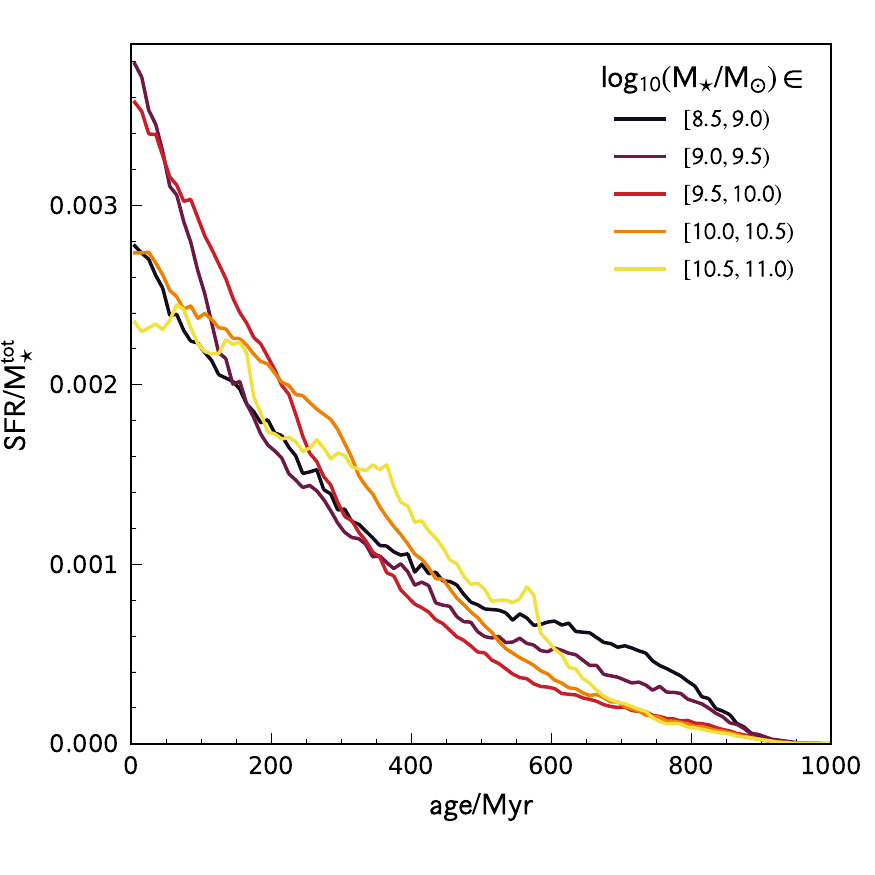}
	\caption{The weighted average (stacked) star formation histories of galaxies at $z=5$ in various stellar mass bins. \label{fig:sf.stacked_sfh}}
\end{figure}

We begin, in Figure \ref{fig:sf.stacked_sfh}, by presenting the average (stacked) star formation histories of galaxies at $z=5$ in bins of stellar mass. With the possible exception of the highest mass-bin, where we see tentative evidence for a plateau, the average star formation history is rapidly increasing. In the subsequent section we quantify the shapes of individual star formation histories using a set of metrics.

\subsection{Metrics}\label{sec:sf.metrics}

We now turn our attention to quantifying the star formation histories of individual galaxies using a variety of metrics. In most cases we present the median of the population alongside both the central 68.4\% ($P_{84.2}-P_{15.8}$) and 95.6\% ranges ($P_{97.8}-P_{2.2}$). The median and these ranges are calculated by applying a weight to each galaxy dependent on its parent re-simulation to ensure the selected regions are representative of the full parent volume.  

\subsubsection{Specific Star Formation Rates}\label{sec:sf.metrics.ssfr}

The most commonly utilised and observationally accessible metric of the SFH is the specific star formation rate (sSFR), the ratio of recent to integrated star formation activity normally expressed as the current star formation rate divided by the stellar mass. In this work we employ the star formation rate averaged over 50 Myr. In \S\ref{sec:appendix.avgsfr} (and specifically Fig. \ref{fig:sim.timescale}) we explore the impact of this assumption finding a relatively small impact ($<0.1$ dex).

\begin{figure*}
	\includegraphics[width=2\columnwidth]{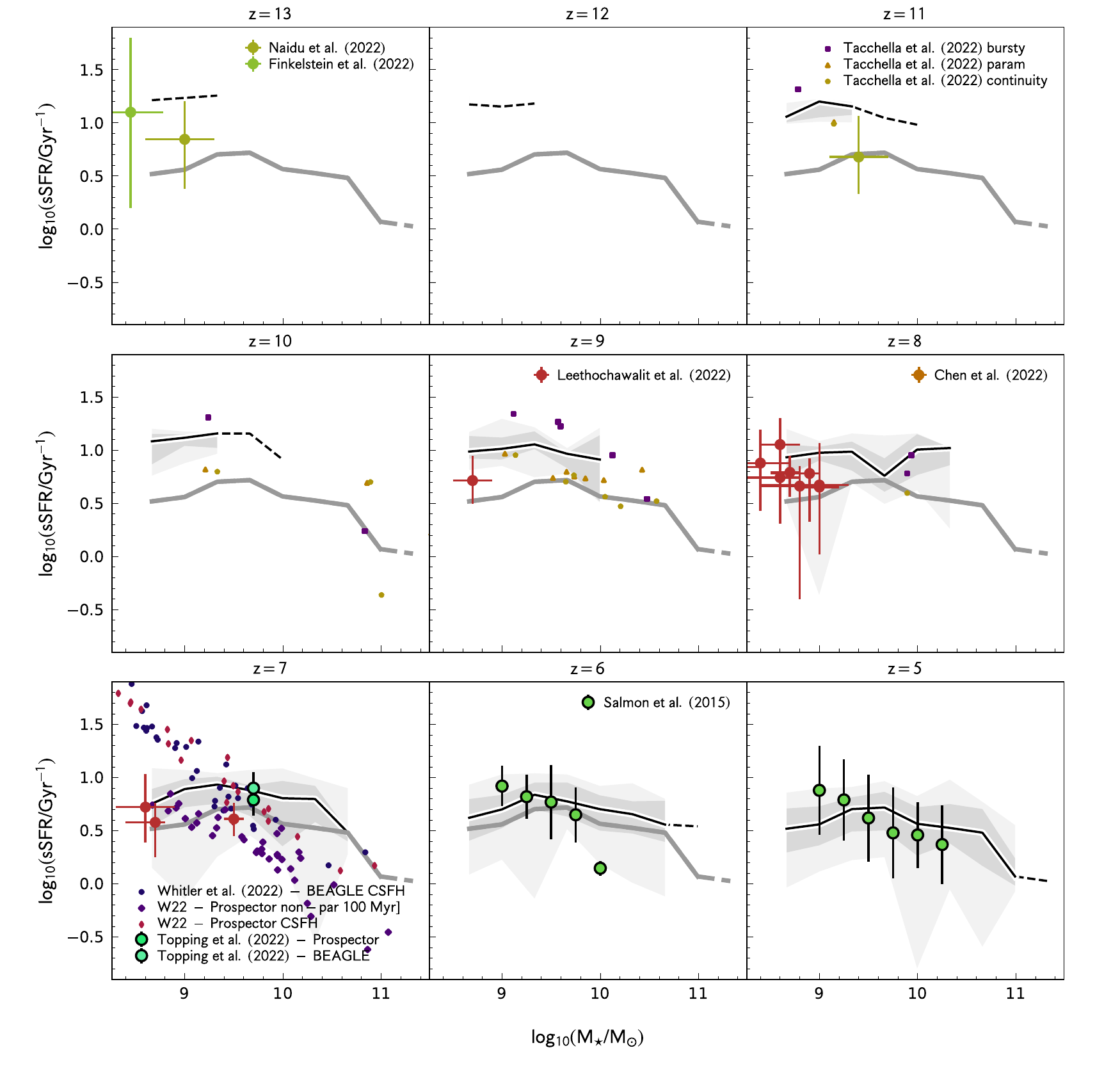}
	\caption{The predicted relationship between the specific star formation rate (sSFR) and stellar mass for integer redshifts $z\in\{5,13\}$. The black solid line denotes the weighted binned median. The two shaded regions show the central 68\% and 95\% ranges. Where the number of galaxies in each bin falls below 10 the solid line is replaced by a dashed line and we no longer present the range. The thick grey line in each panel denotes the median relation at $z=5$. Also shown are various observational constraints including constraints on individual galaxies from \citet{Tacchella22}, \citet{Whitler2022} \citep[based on the sample identified by][]{Endsley21}, \citet{Carnall22}, \citet{Naidu2022}, \citet{Leethochawalit22}, \citet{Finkelstein2022b}, and \citet{Chen2022}, and stacked results from \citet{Salmon2015} and \citet{Topping22}. Uncertainties on the stacked measurements of \citet{Salmon2015} and \citet{Topping22} are the error on the median, not the distribution width. For the \webb\ based results we included the published uncertainties, however for the \spitzer-based results we omit uncertainties for readability.
	\label{fig:sf.ssfr}}
\end{figure*}

\begin{figure}
	\includegraphics[width=\columnwidth]{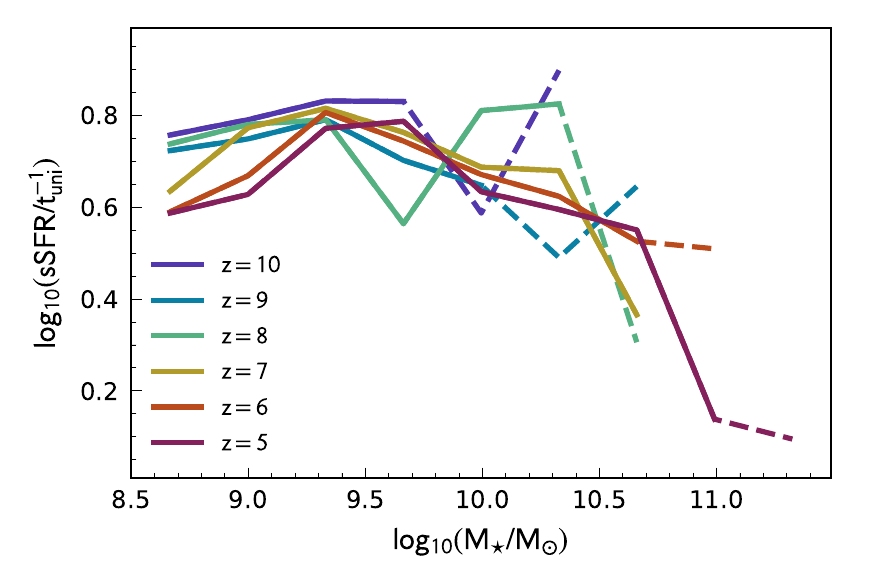}
	\includegraphics[width=\columnwidth]{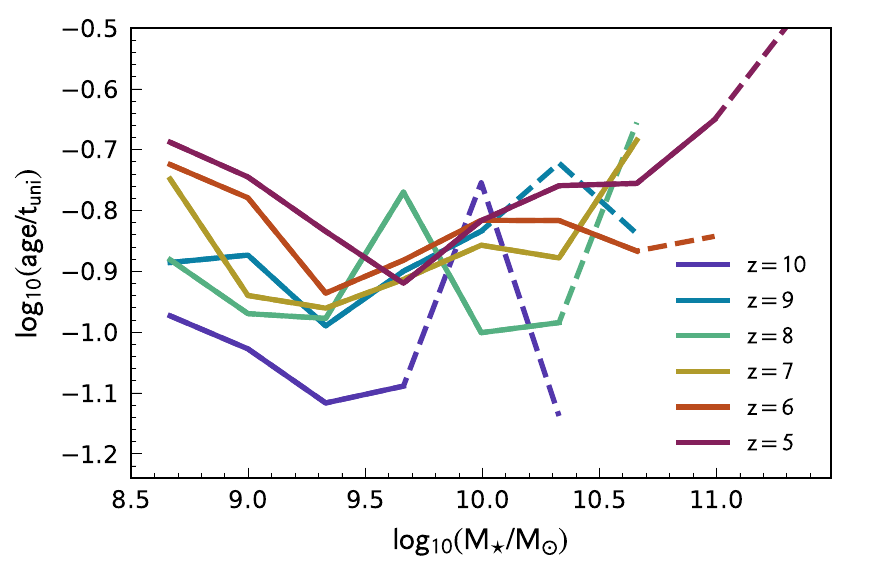}
	\caption{The redshift evolution of specific star formation rate (top) and age (bottom) expressed in terms of the age of the Universe. \label{fig:sf.evo}}
\end{figure}

The resulting relationship between sSFR and stellar mass - also known as the star forming main-sequence - is shown for $z=5-13$ in Figure \ref{fig:sf.ssfr}. This reveals a clear redshift evolution: from $z=13\to 5$ the average sSFR drops by $\sim 0.7$ dex. The evolution of both the specific star formation rate and age (see \S\ref{sec:sf.metrics.age}) is also presented in Figure \ref{fig:sf.evo} but in terms of the age of the Universe. When expressed in this way the redshift evolution is reduced, though some still does remain, at least for the age.

\begin{figure}
	\includegraphics[width=\columnwidth]{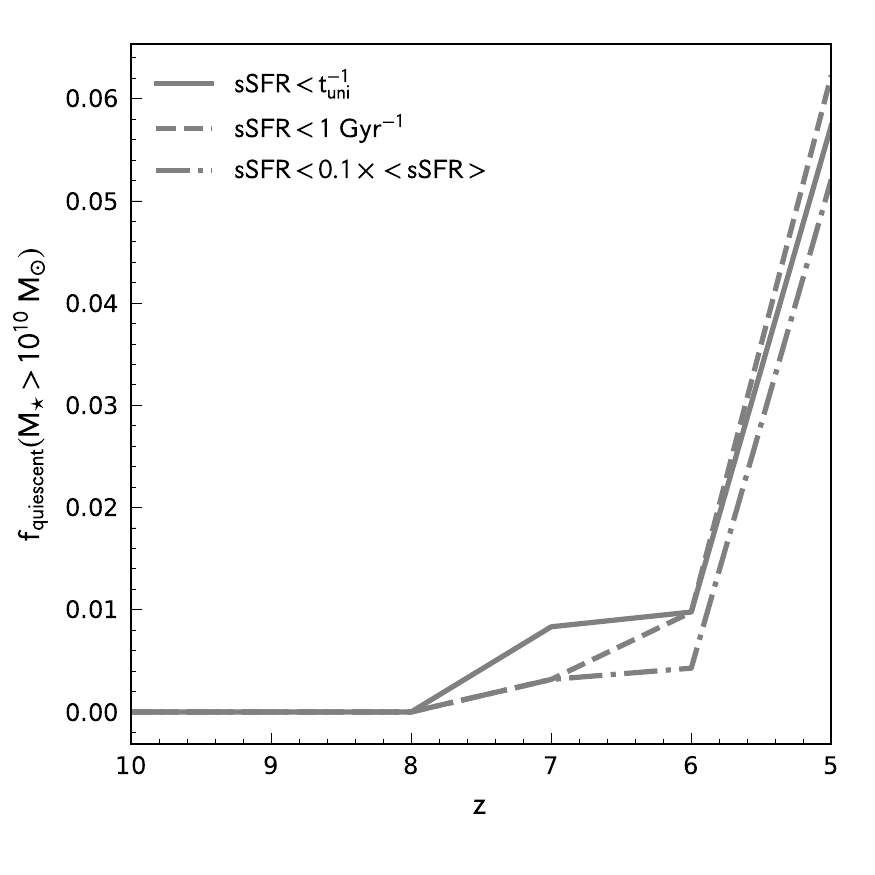}
	\caption{The weighted fraction of massive ($M_{\star}>10^{10}\, {\rm M_{\rm odot}}$) galaxies falling below a specific star formation rate threshold as a function of redshift $z=10\to 5$. The thresholds are the inverse age of the Universe (solid) line, $1\ {\rm Gyr^{-1}}$ (dashed line), and 10\% of the median sSFR (dash-dotted line). \label{fig:sf.quiescent}}
\end{figure}

While Figure \ref{fig:sf.ssfr} reveals significant redshift evolution there is no strong trend with stellar mass, at least at $M_{\star}=10^{8.5-10.5}\, {\rm M_{\odot}}$. At $z=5$, from a peak at $M_{\star}\approx 10^{9.5}\, {\rm M_{\rm \odot}}$ the average specific star formation drops by only $\approx 0.2$ dex by $M_{\star}\approx 10^{10.5}\, {\rm M_{\rm \odot}}$. There is tentative evidence of a sharper drop by $M_{\star}=10^{11}\, {\rm M_{\rm \odot}}$ but \flares\ contains only a handful galaxies at this mass. 

However, while there is no precipitous drop in the average sSFR, there is a sharply increasing fraction of galaxies that have much lower specific star formation rates. In Figure \ref{fig:sf.quiescent} we plot the weighted fraction of massive $M_{\star}>10^{10}\, {\rm M_{\rm \odot}}$ galaxies with specific star formation rates falling below three thresholds: the inverse age of the Universe, $1\ {\rm Gyr^{-1}}$, and 10\% of the median sSFR. Irrespective of the choice of threshold we see that the fraction rapidly increases from $\approx 0$ at $z\ge 8$ to $\sim 0.05$ at $z=5$. The origin of this rapid evolution will be discussed in a companion paper (Lovell \emph{et al.}), though it appears that a reduction in the sSFR is strongly correlated with significant AGN activity. 

\subsubsection{Ages}\label{sec:sf.metrics.age}

Another common metric is the age, though this varies in definition. In this work we define the age as the initial-mass weighted median of the stellar particle ages. Essentially this is the time since the first 50\% of stars were formed. It is important to note however that the term age could also used to describe the mean age; in the context of \flares\ the mean and median ages are similar - and exhibit similar trends - but are not identical (see \S\ref{sec:sf.metrics.moments}). It is also worth noting that an additional literature definition of age is the duration since star formation began \cite[e.g.][]{Laporte21}, sometimes described as maximum age. In the context of simulations however this is expected to be resolution dependent and subject to large uncertainties, and for this reason we do not utilise it in this analysis.

\begin{figure*}
	\includegraphics[width=2\columnwidth]{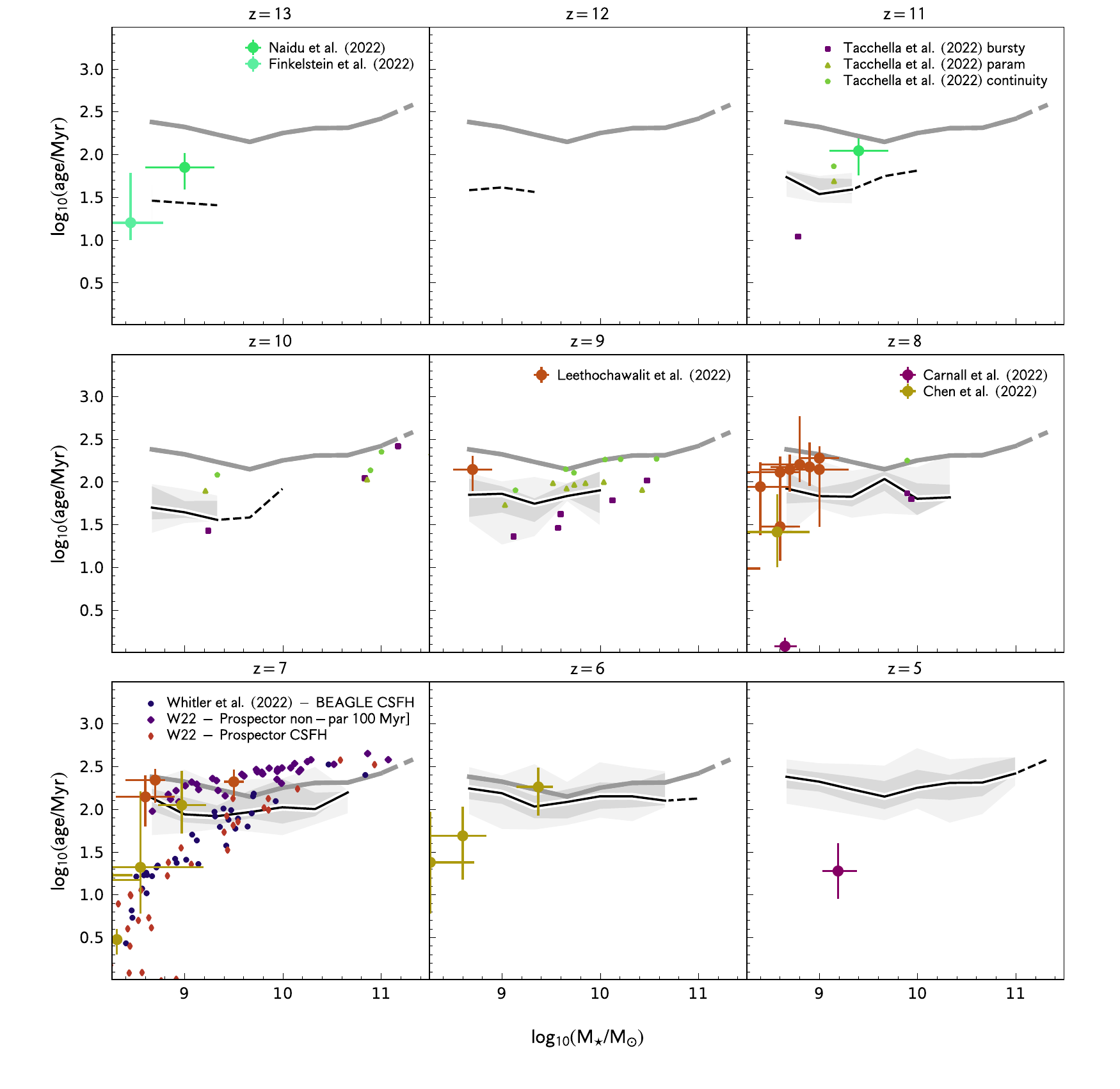}
	\caption{The same as Fig. \ref{fig:sf.ssfr} but for the mass-weighted median age.\label{fig:sf.age}}
\end{figure*}

The relationship between the predicted age and stellar mass is shown in Figure \ref{fig:sf.age}. This relationship largely mirrors the trends seen for the sSFR - stellar mass - redshift plane: average ages decrease with increasing redshift while remaining largely flat with stellar mass.

\footnotetext{The \citet{Finkelstein2022b} source - "Maisie's Galaxy" - has a reported photometric redshift of $z=14.3$, however we include it in our $z=13$ panel.}

\subsubsection{Shape}\label{sec:sf.metrics.shape}

\begin{figure*}
	\includegraphics[width=2\columnwidth]{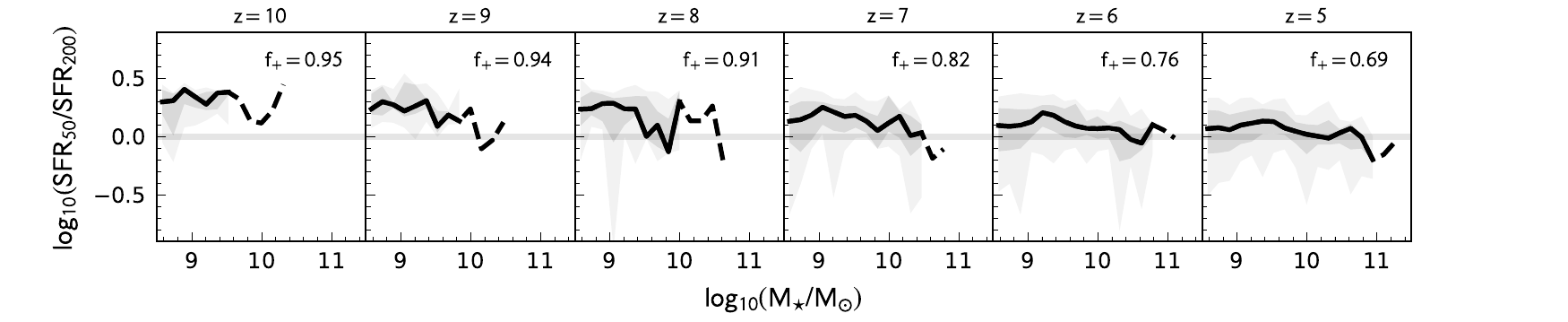}
	\caption{The ratio of the SFR averaged over 50 and 200 million years. The solid horizontal line denotes ${\rm SFR}_{50}/{\rm SFR}_{200}=1$ - galaxies on this line have constant star formation histories. The label denotes the fraction of galaxies that have rising star formation histories measured over this timescale.\label{fig:sf.shape}}
\end{figure*}

While the specific star formation rate gives some indication of the shape of the recent star formation history a clearer picture is revealed by simply comparing the star formation activity averaged over two timescales. We do exactly this in Fig. \ref{fig:sf.shape} where we show the star formation activity measured over the last 50 Myr compared to that measured over 200 Myr. The majority of galaxies at all redshifts have ${\rm SFR}_{50}/{\rm SFR}_{200}>1$ meaning they have star formation histories that are rising over these timescales. However, the fraction of galaxies with rising star formation histories declines from approximately unity at $z=10$ to $\approx 70$\% at $z=5$. Again, there is little trend with stellar mass though the most massive galaxies tend to have a lower fraction of galaxies with still increasing star formation histories.

\subsubsection{Moments}\label{sec:sf.metrics.moments}

\begin{figure*}
	\includegraphics[width=2\columnwidth]{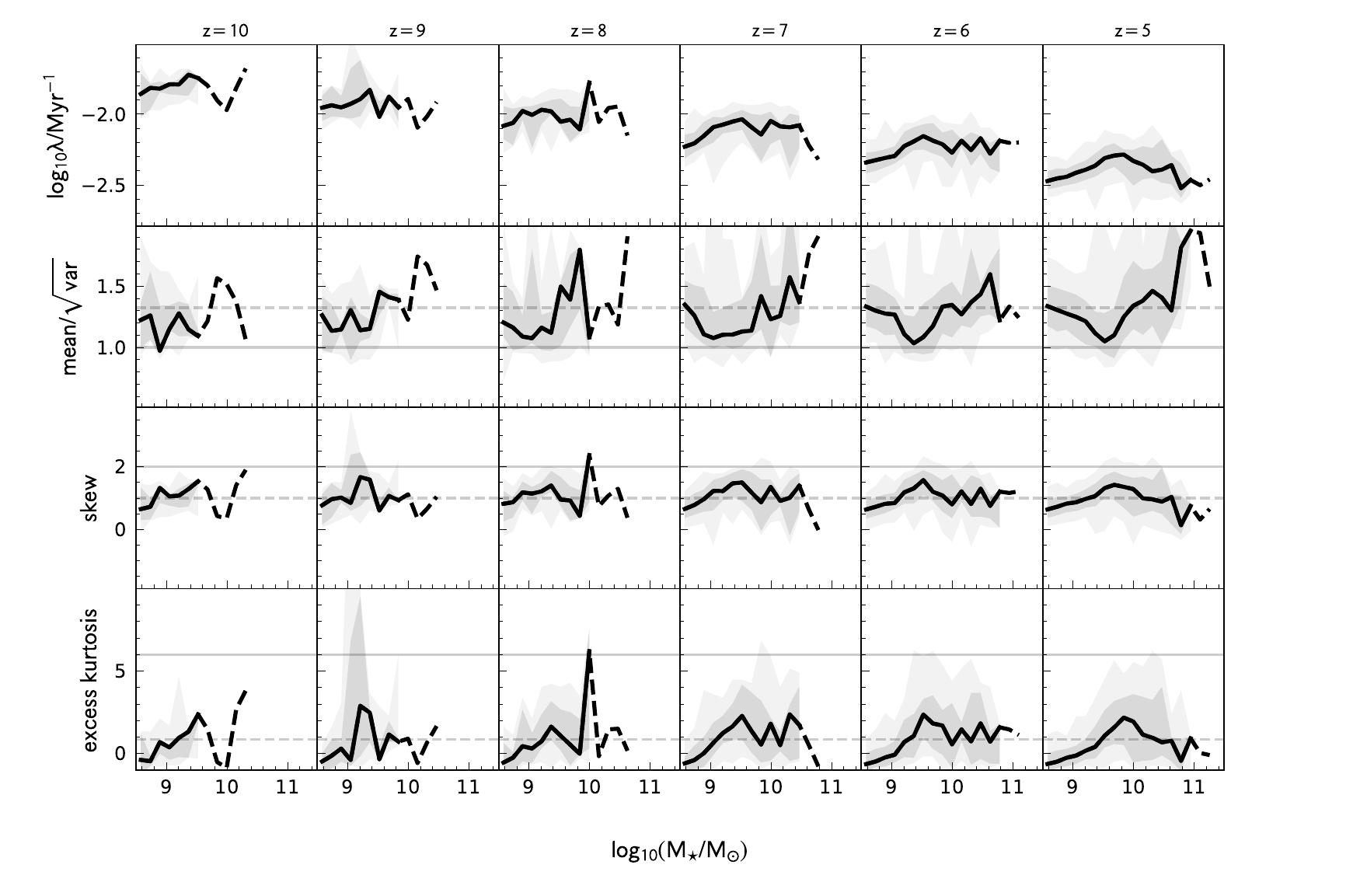}
	\caption{The first four moments of the mass-weighted stellar age distribution for galaxies in \flares\ with $\log_{10}(M_{*}/{\rm M_{\odot}})>8.5$ at $z=10\to 5$. To avoid confusion with the median age we present the inverse of the first moment. For an exponential distribution this is equivalent to the parameter $\lambda$. For the second moment we instead present the quantity ${\rm mean}/\sqrt{{\rm var}}$, which for an exponential distribution is simply unity and for a half-normal is $\approx 1.32$. We also add lines for the expected skew and excess kurtosis of an exponential (solid line) and half-normal distribution (dashed line). \label{fig:sf.moments}}
\end{figure*}

An alternative set of metrics are the moments of the (mass-weighted) stellar age distribution. These are useful as they can be compared directly to various distributions e.g. normal, exponential, and half-normal for which some of the moments have fixed values. The first four moments are presented in Fig. \ref{fig:sf.moments}. Since the first moment is the mean this is similar, but not identical, to our definition of the age (the median). To avoid confusion with the age we present the inverse of the mean and label it $\lambda$. The second moment is the variance; but instead of presenting this directly we present the quantity ${\rm mean}/\sqrt{{\rm var}}$ which has fixed values of unity and $\approx 1.32$ for an exponential and half-normal distribution respectively. This immediately reveals that galaxy star formation histories are not statistically well described by an exponential distribution, particularly at the high mass end for $z<9$. This is also reflected in the third and fourth moments of the distribution, the skew and excess kurtosis respectively; in both cases the measured values lie below those expected (2 and 6 respectively) for a pure exponential distribution. On the other hand the predicted median value of the ${\rm mean}/\sqrt{{\rm var}}$, skew, and excess kurtosis all closely match that expected for a half-normal distribution peaking at the observation epoch suggesting that this provides a useful single-parameter description of galaxy star formation histories in \flares.  

\subsection{Star formation history parameterisation}\label{sec:sf.parameterisation}

\begin{figure*}
	\includegraphics[width=2\columnwidth]{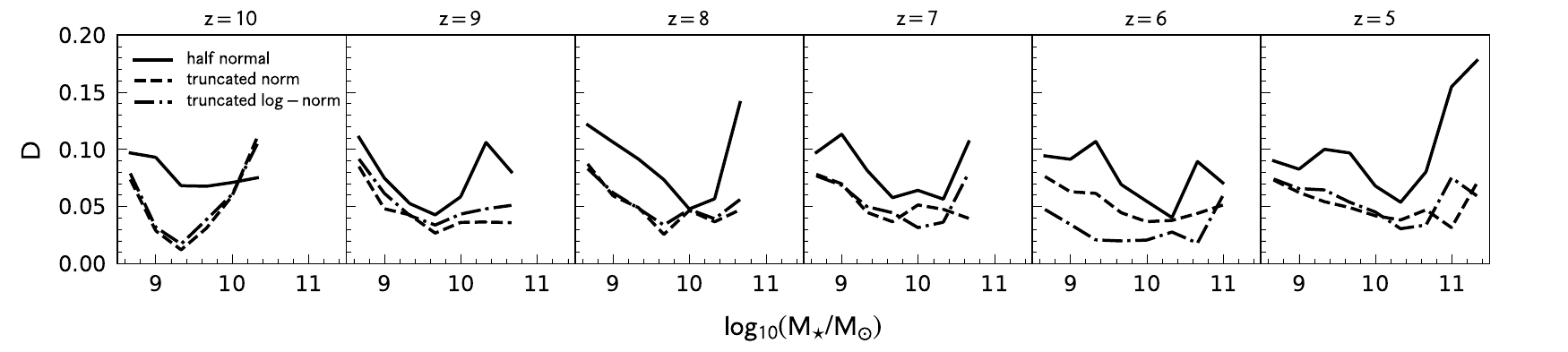}
	\caption{The median value of the Kolmogorov–Smirnov test statistic $D$ for half-normal, truncated normal, and truncated log-normal star formation history. \label{fig:sf.distribution_comparison}}
\end{figure*}

While a half-normal distribution reproduces the skew and variance it assumes we are always observing galaxies at the peak of their star formation histories, something which is clearly not possible. Such a description also fails to account for the increasing numbers of galaxies, revealed in Figure \ref{fig:sf.shape}, which have declining star formation histories.  As an alternative to the single parameter half-normal distribution we also consider the two parameter truncated normal and truncated log-normal distributions,  which have been shown to achieve good fits to simulated and observed SFHs at low redshift \citep{Diemer2017}. We fit the star formation histories of every galaxy by these distributions and computed the Kolmogorov–Smirnov test statistic $D$ with the results presented in Fig. \ref{fig:sf.distribution_comparison}. This analysis reveals that both truncated normal and truncated log-normal distributions yield similar level of improvement over a half-normal distribution. For parametric spectral energy distribution fitting we then advocate one of these distributions as the most suitable parameterisation.

\subsection{Environmental dependence}\label{sec:sf.delta}\label{sec:sf.environment}

\begin{figure*}
	\includegraphics[width=2\columnwidth]{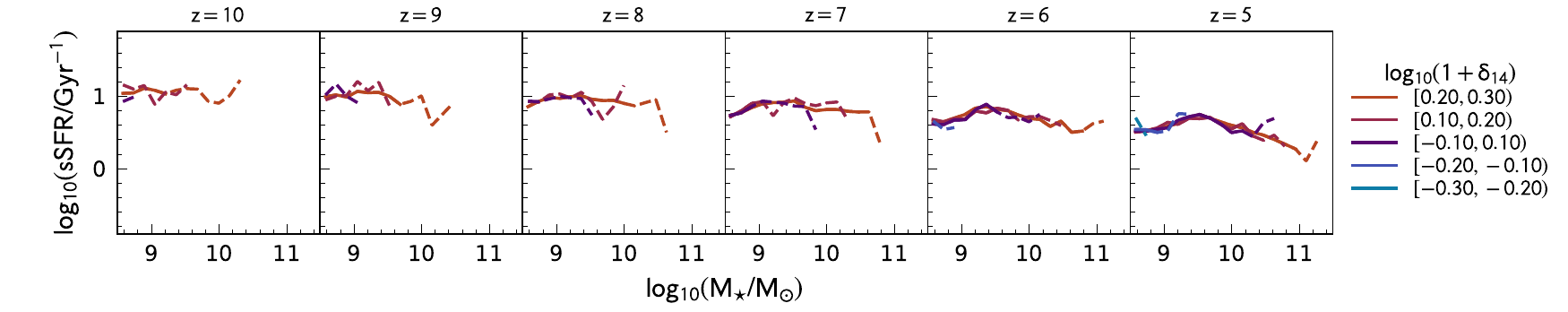}
	\includegraphics[width=2\columnwidth]{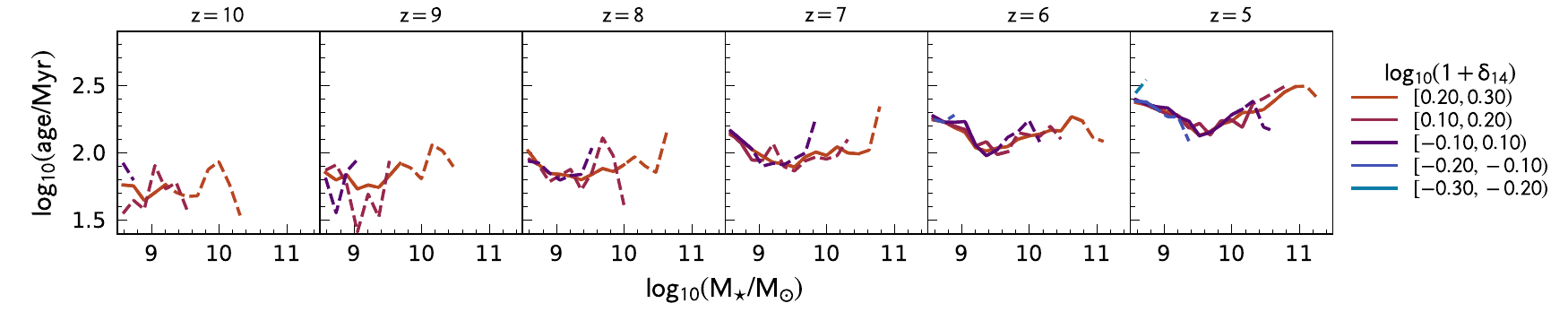}
	\caption{The relationship between the specific star formation rate (top) and age (bottom) as a function of stellar mass as split by the density contrast of the simulation. \label{fig:sf.delta}}
\end{figure*}

A unique feature in  \flares\ is the re-simulation of a wide range of environments from $\log_{10}(\delta_{15})=-0.3\to 0.3$. This allows us to study the environmental dependence of the shape of the SFH. In Fig. \ref{fig:sf.delta} we present the relationship between stellar mass and sSFR and age as a function of galaxy environment. Matching that found in \citet{FLARES-I} we see no environmental dependence of these properties. However, it is important to note that this is not to say that environment doesn't play a strong role in galaxy formation at high-redshift: there is a strong environmental bias, such that under-dense regions contain proportionally many fewer galaxies, and vice versa. In addition there is an implicit dependence of galaxy mass on environment: only the most over-dense regions produce massive galaxies at high-redshift.

\subsection{Comparison with observational constraints}\label{sec:sf.observations}

At the time of writing we are now seeing the first constraints on the star formation histories of galaxies from \webb\ \citep[][]{Carnall22, Naidu2022, Leethochawalit22, Finkelstein2022b, Chen2022}. Prior to \webb\ however, it was possible, by combining ground based or \hubble\ optical and near-IR observations with \spitzer, to constrain the rest-frame UV - optical emission of $z>5$ galaxies. Alongside the \flares\ predictions in Figs \ref{fig:sf.ssfr} and \ref{fig:sf.age} we show current observational constraints on the specific star formation rate and age, including early results from \webb\ and previous results from \hubble\ + \spitzer. However, it is important to note that these observed samples are unlikely to be complete in $M_{\star}$, and in many cases will be biased to high specific star formation rates and low ages since these lead to higher luminosities. Since all of these samples are rest-frame UV selected it is also possible they are missing heavily dust obscured systems and/or systems with strongly quenched star formation histories.

\citet{Salmon2015} studied the evolution and slope of the relationship between star formation activity and stellar mass in a sample of galaxies $3.5\le z\le 6.5$ using multi-wavelength photometry in GOODS-S from the Cosmic Assembly Near-infrared Deep Extragalactic Legacy Survey \citep[CANDELS,][]{CANDELS, Koekemoer2011} and \spitzer\ Extended Deep Survey \citep{SED}. As shown in Figure \ref{fig:sf.ssfr} the individual binned median values in most cases match our predictions within the uncertainties. However, while our results show little variation with stellar mass these tend to decrease as a function of stellar mass. 

\citet{Endsley21} utilised a selection combining narrow-band and overlapping broadband filters yielding precise photometric redshifts, essential for constraining the contribution of strong [O\textsc{iii}] and H$\beta$ line emission. Using this sample \citet{Whitler2022} measured ages and specific star formation rates utilising both the \texttt{Prospector} \citep{Leja2017, Prospector} and \texttt{BEAGLE} \citep{BEAGLE} spectral energy distribution (SED) fitting codes. For the \texttt{BEAGLE} analysis a constant SFH model while the \texttt{Prospector} analysis explored both constant and non-parametric star formation histories. As seen in Fig. \ref{fig:sf.ssfr} both constant SFH implementations yielded specific star formation rates at $M_{\star}\sim 10^{9.5}\ {\rm M_{\odot}}$ similar to the \flares\ predictions. However, both produced steeply declining specific star formation rates as a function of stellar mass, contrary to the mostly flat relationship predicted by \flares. Specific star formation rates measured using \texttt{Prospector} assuming non-parametric SFHs yielded a flatter, but still declining, relationship and lower overall normalisation, falling below the \flares\ predictions. The observed trends in the age stellar mass relation (Fig. \ref{fig:sf.age})  mirror the trends in specific star formation rate: the average ages at $M_{\star}\approx 10^{9.5}\ {\rm M_{\odot}}$ are similar to the \flares\ predictions but diverge at lower and higher masses. The non-parametric analysis resulted in larger ages and a much shallower trend with stellar mass, albeit with ages offset to larger values than predicted by \flares. Beyond a model issue, one possibility for the difference in the slope of observations relative to the \flares\ predictions is the impact of dust. While the SED modelling takes account of dust, it is possible the obscured star formation in the most massive galaxies has been underestimated resulting in systematically low stellar masses. 

\citet{Tacchella22} studied the stellar populations of a sample of 11 bright galaxies at $z=9-11$ selected from the CANDELS fields \citep{Finkelstein2022}. These galaxies were analysed using the \texttt{Prospector} using a range of non-parametric and parametric star formation history priors. Our predictions are bracketed by the \citet{Tacchella22} measurements with those assuming a continuity or parametric prior falling below our predictions and those with a bursty prior above. In addition to specific star formation rates \citet{Tacchella22} also measured the ages of their sample of galaxies. Unsurprisingly these yield a similar trend to what was found for the specific star formation rates: \flares\ predictions are bracketed by measurements made assuming various priors.

\citet{Topping22} measured the specific star formation rates of specific of 40 UV-bright galaxies at $z\sim7-8$ combining far-IR continuum and [C\textsc{ii}] constraints from ALMA with observations from \hubble\ and \spitzer. \citet{Topping22} also utilised both \texttt{Prospector} and \texttt{BEAGLE}, though recovered similar average specific star formation rates. Both measurements are consistent with \flares\ within their uncertainties. 

\citet{Carnall22} studied 5 spectroscopically confirmed ($z=5-9$) lensed (lensing factor $=1.5-10$) in the SMACS0723 Early Release Observation (ERO) NIRCam imaging. \citet{Carnall22} fit the observed NIRCam F090W, F150W, F200W, F277W, F356W, and F444W fluxes using the \texttt{Bagpipes} SED fitting code \citep{Bagpipes} obtaining very young ages: 4 objects have mean stellar ages $\le 2$ Myr and one has $\approx 20$ Myr. Because these objects are both relatively faint and lensed, 3/5 have stellar masses fall outside our predictions. However, the two that do fall within our range are, unlike most other observational measurements, discrepant with our predictions. This may suggest \flares\ fails to fully capture the full range of possible star formation histories or reflects an observational/modelling issue.

\citet{Naidu2022} identified two bright $z>10$ candidates in the GLASS Early Release Science (ERS) imaging data based on a search of both GLASS and CEERS NIRCam observations. The SEDs of these objects were then fit with \texttt{Prospector} code assuming a continuity prior. The resulting specific star formation rates fall slightly below our predictions at their respective redshifts, though are consistent within the observational uncertainties and predicted scatter.

\citet{Leethochawalit22} identified and studied a sample of 14 galaxies at $7<z<9$ in GLASS Early Release Science (ERS) NIRCam imaging and measured their star formation histories using \texttt{Bagpipes}. Most of the inferred ages and specific star formation rates are in generally good agreement with our predictions, albeit with a slight (0.2 dex) offset to higher ages and lower specific star formation rates. 


\citet{Finkelstein2022b} identified a sole $z>13$ candidate, "Maisie's Galaxy" at $z\approx 14.3$ in CEERS NIRCam imaging. \citet{Finkelstein2022b} measure stellar masses, ages, and specific star formation rates using \texttt{Prospector} finding an age of $16^{+45}_{-5}\ {\rm Myr}$, consistent with our predictions. 

\citet{Chen2022} study a sample of 12 galaxies at $6<z<8$ using CEERS NIRCam imaging. These sources were previously identified from \hubble\ observations of the EGS field. \citet{Chen2022} measured the star formation histories of these galaxies using \texttt{Beagle}. Where the stellar masses of the \citet{Chen2022} sample overlap with our predictions there is good agreement, however, the galaxies at lower stellar masses have smaller ages than expected from an extrapolation of the \flares\ trend. Like the \citet{Carnall22} study this may mean that \flares\ fails to predict the full range of possible star formation history scenarios. 

These comparisons reveal a mixed picture, with some tentative evidence for differences in the slope and normalisation of the specific star formation rate - stellar mass relation. Beyond a model issue one observational solution is potentially the effect of dust; all of these observational constraints are based on UV selected sources and, except for the case of \citet{Topping22}, utilise only rest-frame UV and optical observations. This may not only mean that dusty intensely star forming galaxies are missing but that the contribution from dust obscured star formation in the most massive galaxies is underestimated. Furthermore, it is clear from both the \citet{Tacchella22} and \citet{Whitler2022} analyses, in addition to other \citep[e.g][]{Leja2019, Carnall2019} studies that these constraints are strongly sensitive to modelling assumptions, such as the choice (or not) of star formation history parameterisation. On the modelling side, the relatively low resolution of \flares\ limits the low-mass galaxy populations we can model at very high redshifts $z > 10$. The \eagle\ model also ignores the effects of radiative feedback and reionisation, which may have a particularly large effect on low mass halos. It is promising, however, that a model calibrated at $z = 0$ shows such promising agreement with a range of observational constraints in this high redshift domain.

\section{Metal Enrichment}\label{sec:Z}

We now turn out attention to the predictions for the mass-weighted \emph{stellar} metallicity $Z_{\star}$ of galaxies. In this work we focus on stellar metallicities $Z_{\star}$ deferring predictions for gas-phase metallicities, including predictions for observable line ratios etc. to a future work.

\subsection{Earlier work and limitations}

\begin{figure}
	\includegraphics[width=\columnwidth]{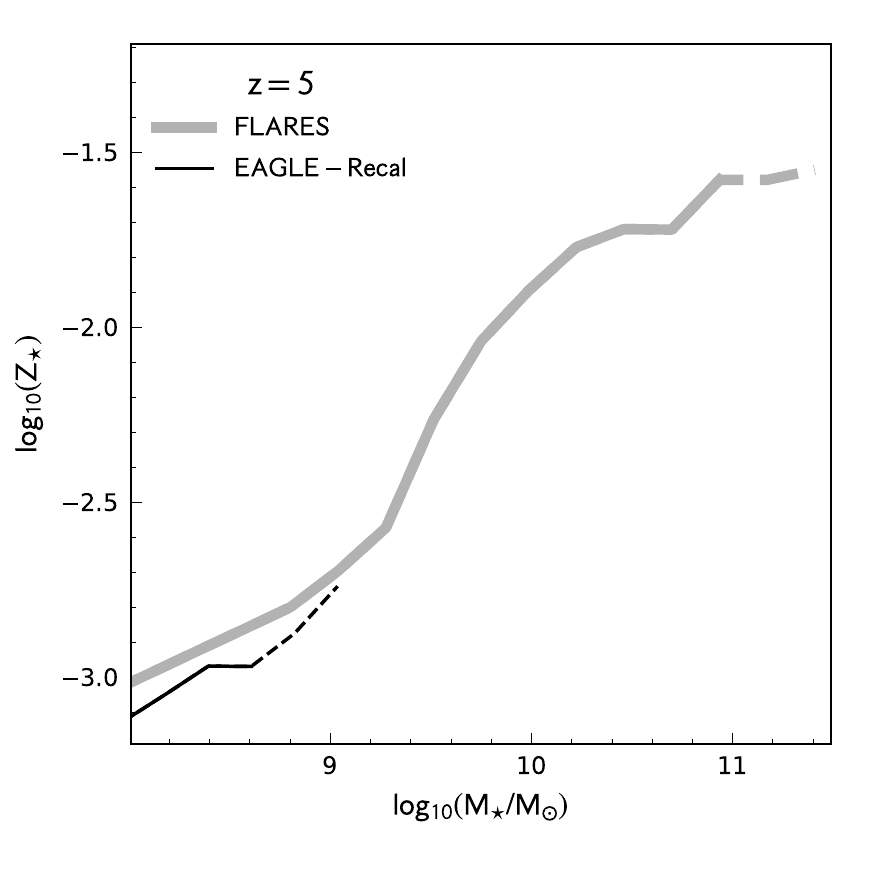}
	\caption{Comparison between the stellar mass-metallicity relation predicted by \flares\ and the \eagle-ReCal model at $z=5$. \label{fig:Z.MZR_Recal}}
\end{figure}

The stellar mass - stellar metallicity relation predicted by the \eagle\ model at $z=0$ was presented in \S6.3 of \citet{EAGLE}, with this analysis subsequently extended by \citet{DeRossi2017} to $z=3$. In particular \citet{EAGLE} explored predictions for the reference (Ref), AGNdT9, and higher-resolution recal (ReCal) model. While there is good agreement for high-mass ($M_{\star}>10^{10}\ {\rm M_{\odot}}$) galaxies, stellar metallicities diverge at lower stellar masses, with the discrepancy maximised at $M_{\star}=10^{8-9}\ {\rm M_{\odot}}$ such that galaxies in the Ref/AGNdT9 variant simulations have stellar metallicities $0.2-0.3$ dex larger than those in the ReCal simulation. To understand whether this offset still exists at $z=5$, in  Figure \ref{fig:Z.MZR_Recal} we compare the ReCal and \flares\ stellar mass-metallicity relations at $z=5$. While the small volume of the ReCal simulation compared to \flares\ precludes a comparison at $M_{\star}>10^{9}\ {\rm M_{\odot}}$, at $M_{\star}=10^{8-9}\ {\rm M_{\odot}}$ the offset is only $\approx 0.1$ dex, giving some confidence that our predictions at higher masses will be robust. 

\subsection{Mass - metallicity relation}\label{sec:Z.MZR}

\begin{figure*}
	\includegraphics[width=2\columnwidth]{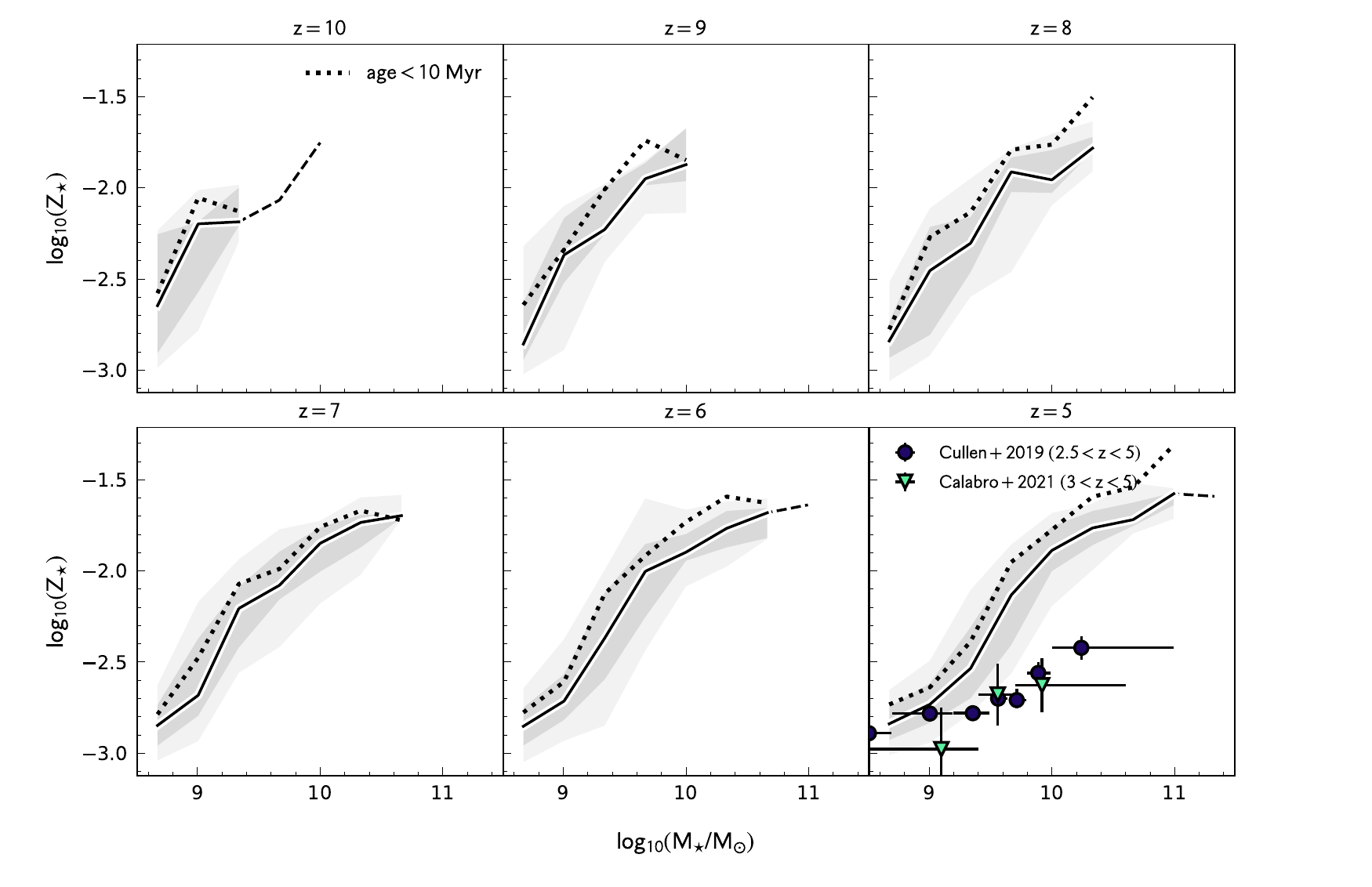}
	\caption{The same as Fig. \ref{fig:sf.ssfr} but for the total stellar metallicity. The dotted line shows the median stellar metallicity of only young ($<10$ Myr) star particles. In the $z=5$ panel we show observational constraints, at $2.5<z<5$, from  \citet{Cullen19} and \citet{Calabro21}. \label{fig:Z.MZR}}
\end{figure*}

We next present the \flares\ stellar mass-metallicity relation at $z=5$--10 in Figure \ref{fig:Z.MZR}. Unlike the age, stellar metallicities exhibit a strong dependence on stellar mass, increasing by a factor of 10 over the mass range $10^{9}\to 10^{11}\ {\rm M_{\odot}}$. Figure \ref{fig:Z.MZR} also includes the median stellar metallicity for the youngest (age $<10$ Myr) star particles within each galaxy. The metallicity of young stars follows the same trend with stellar mass as the general stellar mass-metallicity relation but is offset to higher metallicity by $\approx 0.1$--0.2 dex, reflecting the correlation between metallicity and age explored in more detail in \S\ref{sec:Z.age}.

\subsubsection{$\alpha$-enhancement}

\begin{figure}
	\includegraphics[width=\columnwidth]{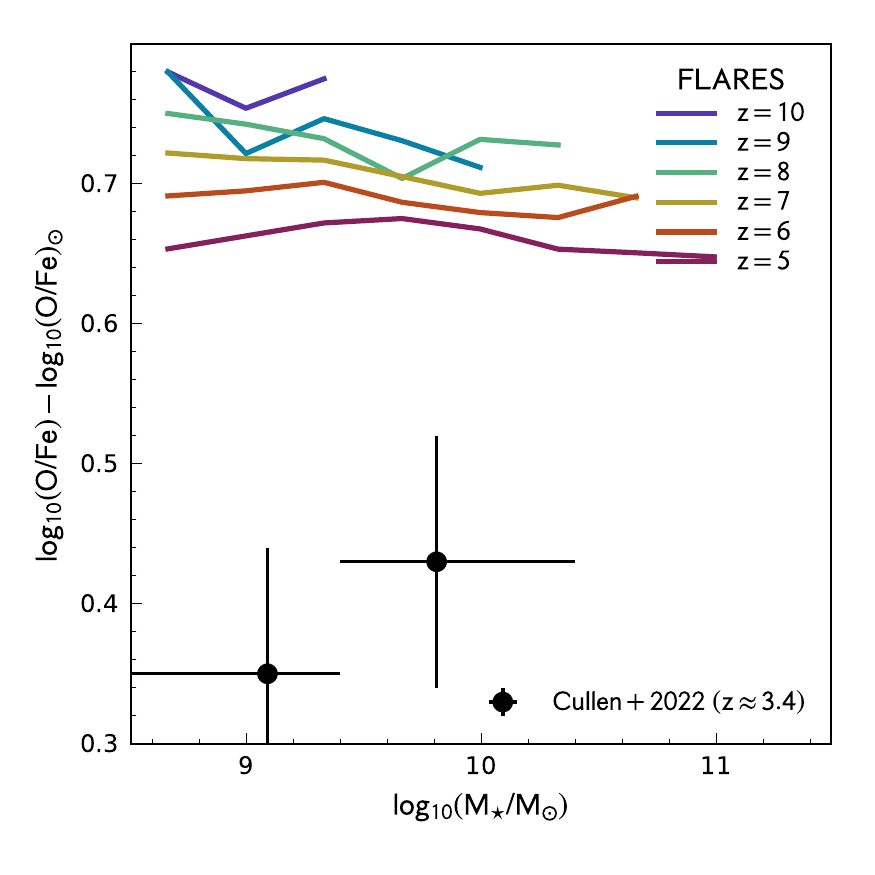}
	\caption{The predicted redshift evolution of $\alpha$-enhancement of galaxies in \flares. Also included are recent constraints from \citet{Cullen21} at $z\approx 3.4$. \label{fig:Z.OFe}}
\end{figure}

In young, star forming galaxies, such as those at high-redshift, the dominant mode of chemical enrichment will be via type II (core-collapse) supernovae (SNII). Consequently early galaxies should be "$\alpha$-enhanced" relative to the galaxies present in the later Universe. In Figure \ref{fig:Z.OFe} we show the $\alpha$-enhancement, quantified as [O/Fe]$=\log_{10}({\rm O/Fe})-\log_{10}({\rm O/Fe})_{\odot}$, as a function of stellar mass and redshift. At $z=10$ we predict [O/Fe]$\approx 0.75$, falling to $\approx 0.65$ at $z=5$. The relationship with stellar mass is mostly flat, mirroring the trends seen in Section \ref{sec:sf} for the age and specific star formation rate. 

\subsubsection{Comparison to observational constraints}

\begin{figure}
	\includegraphics[width=\columnwidth]{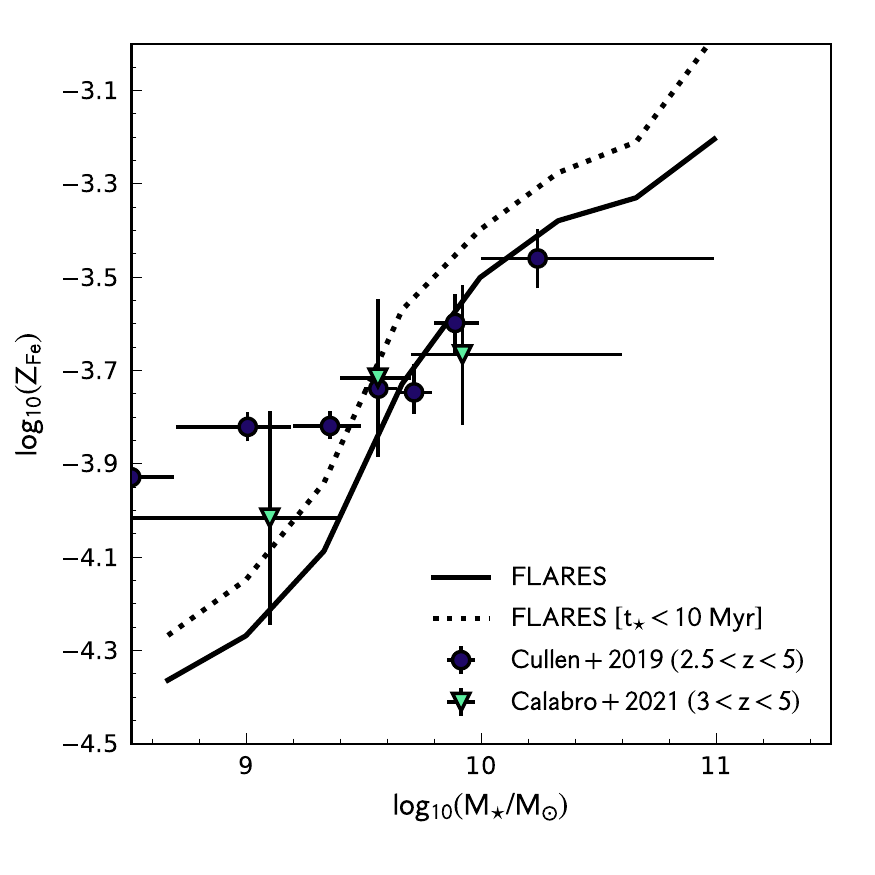}
	\caption{The predicted mass fraction of iron in stars ($Z_{\star, \rm Fe}$) as a function of stellar mass from \flares\ and the observations of \citet{Cullen19} and \citet{Calabro21}. The dotted line shows only the mass fraction of only young ($<10$ Myr) star particles.  \label{fig:Z.MZR_Fe}}
\end{figure}

At present, most galaxy metallicity constraints come from observations of optical emission lines. Integrated across galaxies these are sensitive to the gas-phase oxygen abundance (O/H)$_{g}$. Prior to \webb, measurements were limited to $z\approx 4$ \citep[e.g.][]{Sanders21} however, thanks to \webb's near-infrared spectroscopic capabilities, the first constraints are emerging to $z\sim 8$ \citep[e.g][]{Trump2022}. It is however possible to constrain stellar metallicities from rest-frame observations of the UV continuum where features sensitive to the stellar photospheric iron abundance (Fe/H) exist \citep{Leitherer2010}. Recent efforts \citep[e.g][]{Cullen19, Calabro21} have pushed these measurements to $z>2$ using deep ground-based near-infrared spectroscopy. Constraints on the stellar mass - metallicity relation from \citet{Cullen19} and \citet{Calabro21} are presented in Figure \ref{fig:Z.MZR} alongside our predictions. At $M_{\star}\sim 10^{10}\ {\rm M_{\odot}}$ our predictions are up-to $5\times$ higher than both \citet{Cullen19} and \citet{Calabro21}. However, both \citet{Cullen19} and \citet{Calabro21} are predominantly sensitive to the iron abundance (Fe/H), not the total stellar metallicity. In both cases the total stellar metallicity was inferred by extrapolation using a solar abundance pattern. As noted above however, due to the relative lack of enrichment from Type Ia supernovae, our abundances are extremely $\alpha$-enhanced. If we compare the iron mass fractions directly - using the solar abundances of \citet{Asplund2009} to convert the \citet{Cullen19} and \citet{Calabro21} measurements to an iron mass fraction - we find greatly improved agreement with the \flares\ predictions, at least at the high-mass end (see Figure~\ref{fig:Z.MZR_Fe}). In Figure \ref{fig:Z.OFe} we also compare our $\alpha$-enhancement predictions with \citet{Cullen21} who combined stellar and gas-phase metallicity measurements of the same galaxies at $z\approx 3.4$ to constrain [Fe/O]. These results are $\approx 0.25$ dex below our predictions, though this is likely to be at least partially reconcilable with subsequent redshift evolution. 

With \webb\ now beginning to collect deep near-infrared spectroscopy of high-redshift galaxies observation constraints on both $Z_{\star}$ and $Z_{g}$ will soon dramatically improve. This in turn will begin to provide critical constraints to galaxy formation models.

\subsubsection{Environmental dependence}\label{sec:Z.MZR.delta}

\begin{figure*}
	\includegraphics[width=2\columnwidth]{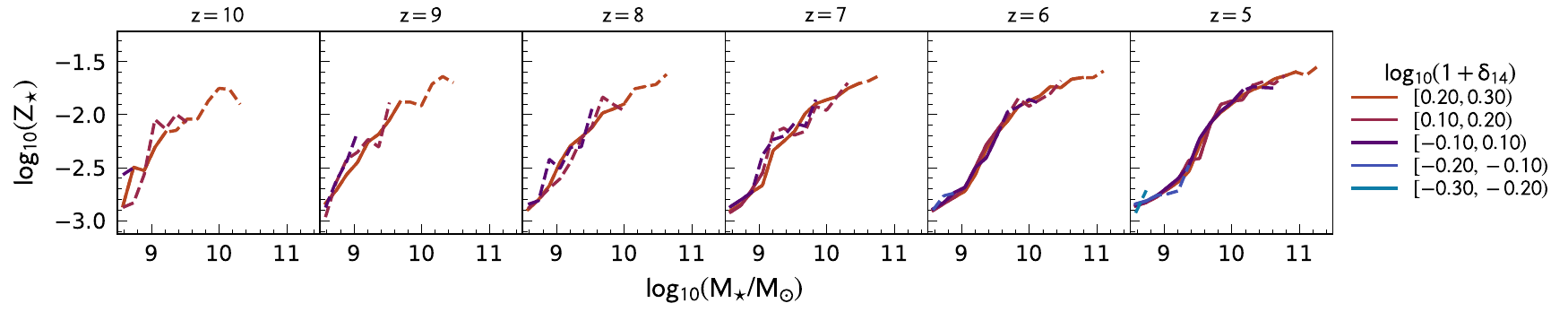}
	\caption{The same as Fig. \ref{fig:sf.delta} but for the metallicity. \label{fig:Z.delta}}
\end{figure*}

\begin{figure}
	\includegraphics[width=\columnwidth]{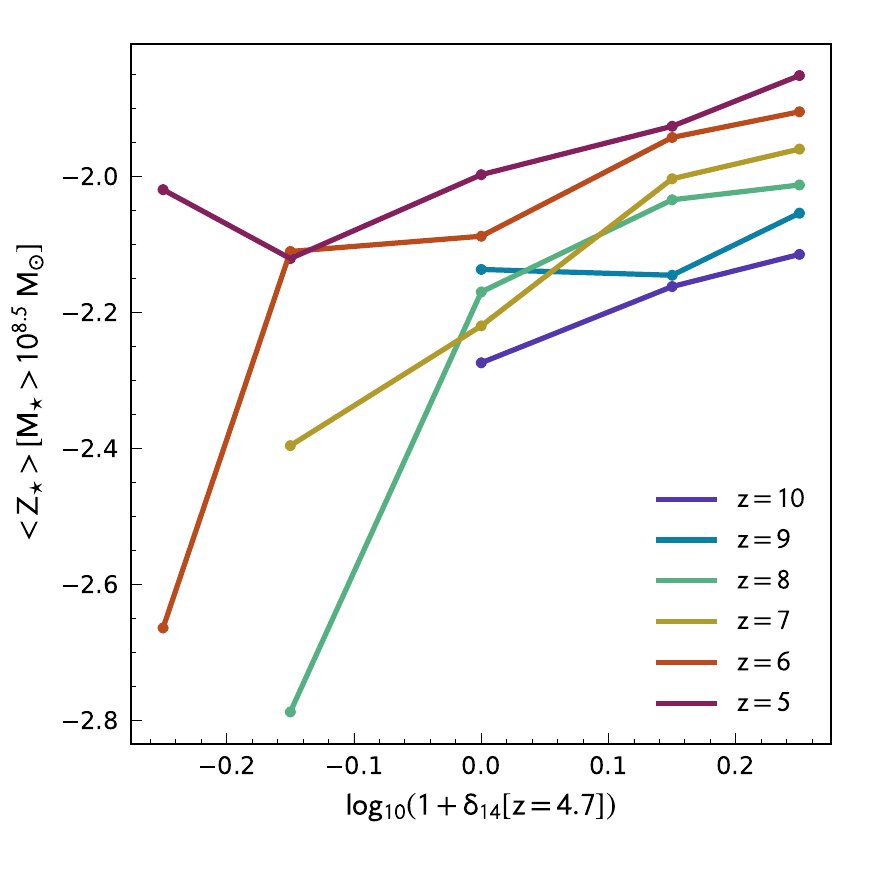}
	\caption{The average mass-weighted metallicity of galaxies with $M_{\star}>10^{8.5}\ {\rm M_{\odot}}$ as a function of environment. \label{fig:Z.delta_total}}
\end{figure}

As demonstrated in Figure \ref{fig:Z.delta} and as seen for the age and specific star formation rate (see \S\ref{sec:sf.delta}, Figure \ref{fig:sf.delta}) we see no evidence of an environmental dependence on the stellar mass - metallicity relationship. This is not to say however that higher-density environments are not more metal enriched, as demonstrated in Figure \ref{fig:Z.delta_total} they are. This Figure shows a clear trend such that the average stellar metallicity is higher in more dense environments. This Figure also shows  clear redshift evolution such that average metallicity increases. The reconciliation of this and Figure \ref{fig:Z.delta} is that higher-density environments, and lower-redshifts, contain a larger fraction of their mass in more massive, metal-rich galaxies. 

\subsection{Correlation with age}\label{sec:Z.age}

\begin{figure*}
	\includegraphics[width=2\columnwidth]{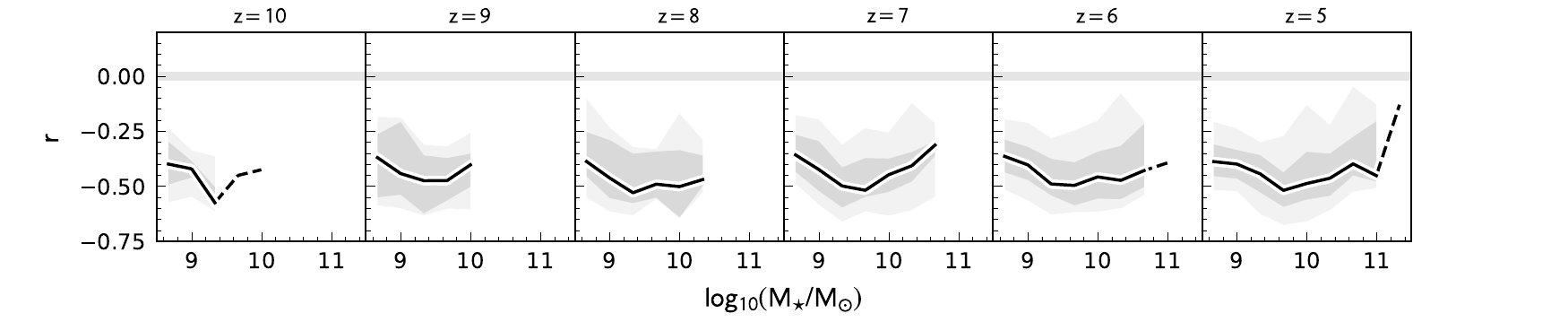}
	\caption{The results of linear fitting to the age and $\log_{10}(Z_{\star})$ of star particles in each galaxy. The top row shows the best-fit slope, the middle row the age$=0$ intercept, and the final row the Pearson correlation coefficient $r$. \label{fig:Z.age}}
\end{figure*}

As preceding generations stars return enriched material to the interstellar medium, subsequent generations are, in general, expected to become increasingly enriched with metals. Consequently we should expect a negative correlation between the stellar age and metallicity. To test this we calculate the Pearson correlation coefficient $r$, finding (see Fig. \ref{fig:Z.age}) typical values of $\approx -0.5$ suggesting a weak negative correlation.

\subsection{Distribution}\label{sec:Z.dist}

\begin{figure}
	\includegraphics[width=\columnwidth]{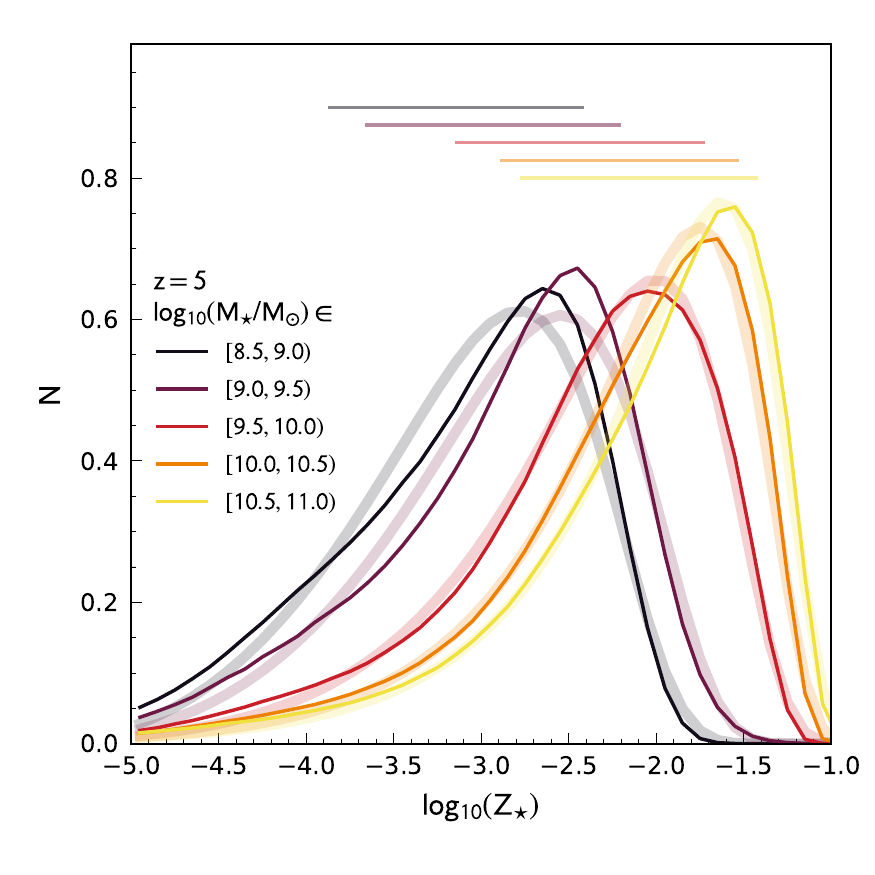}
	\caption{The distribution of stellar metallicities for stacks of galaxies in stellar mass. The horizontal lines denote the central 68.4\% ($P_{84.2}-P_{15.8}$) range of each distribution. The faint thick lines are parametric fits to predicted distribution and assume the (two parameter) exponential power distribution. \label{fig:Z.stacked_Z}}
\end{figure}

\begin{figure*}
	\includegraphics[width=2\columnwidth]{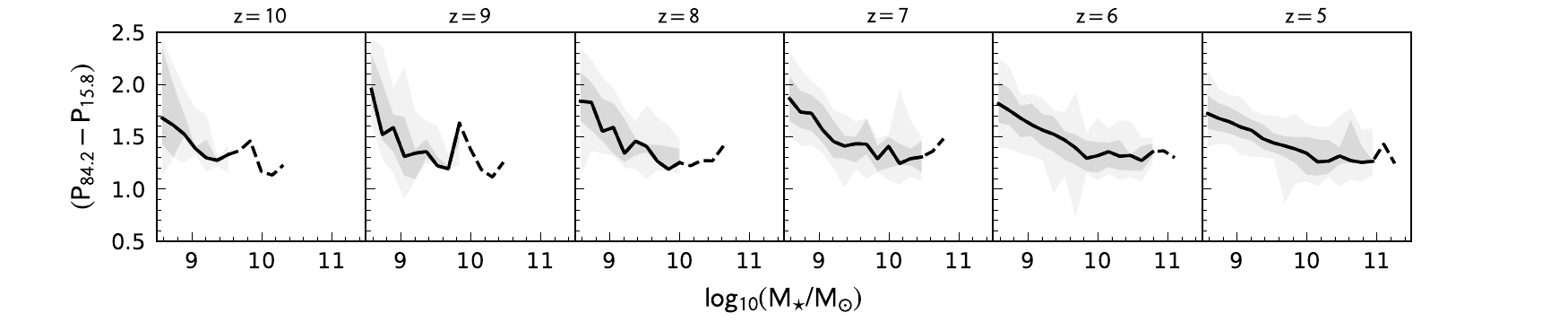}
	\caption{The same as Fig. \ref{fig:sf.ssfr} but for the central 68.4\% range of stellar metallicities. \label{fig:Z.range}}
\end{figure*}

As with ages, stellar populations in galaxies will have a range of metallicities. Figure \ref{fig:Z.stacked_Z} shows the initial mass weighted distribution of stellar metallicities for galaxies stacked by galactic stellar mass. This reveals that galaxies have a broad predicted range of metallicities with the central 68.4\% range of $Z_{\star}$ extending over more than 1 dex. This is also seen for individual galaxies as shown in Figure \ref{fig:Z.range} where the average range is 1-1.5 dex. This figure also reveals a trend with stellar mass such that the most massive galaxies have, on average, higher and narrower metallicity distributions. 

Figure \ref{fig:Z.stacked_Z} also reveals that the distribution of metallicities is asymmetric with a long tail to low metallicity. While this distribution can be fit adequately by a range of standard distributions, none have been found to give consistently good ($D<0.1$) results across the full range of stellar mass stacks and redshifts. Figure \ref{fig:Z.stacked_Z} presents fits assuming the two parameter exponential power distribution\footnote{\url{https://docs.scipy.org/doc/scipy/reference/generated/scipy.stats.exponpow.html}}; while this yields good results in the larger stellar mass bins, the fit becomes progressively worse for the smaller stellar mass bins.  

The fact that individual galaxies have stellar metallicities spanning a wide range, alongside a correlation with age, will have implications for the extent to which it is possible to constrain metallicities and other physical properties from observations using spectral energy distribution fitting. While some codes allow the metallicity to evolve, either through a parametric form \citep[e.g][]{ProSpect}, or in independent bins \citep[e.g][]{VESPA, Prospector}, none that we are aware of explicitly allows the modelling of metallicity distributions, which may lead to significant biases.


\section{Conclusions}\label{sec:conc}

In this work we have presented predictions for the star formation and metal enrichment histories of galaxies at $z=5-10$ from the \flares: First Light And Reionisation Epoch Simulations project. Our main conclusions are as follows:

\begin{itemize}

    \item We find that specific star formation rates strongly evolve with redshift and exhibit a weak negative trend with stellar mass at $M_{\star}>10^{9}\ {\rm M_{\odot}}$. These predictions are in broadly good agreement with many current observational constraints, including early results from \webb. However, there is a tentative evidence of a difference in the slope of the relation and larger scatter present in the observations. 
    
    \item As redshift decreases, an increasing fraction of massive ($M_{\star}>10^{10}\ {\rm M_{\odot}}$) galaxies have relatively low specific star formation rates. This is attributed to AGN feedback and is the focus of a companion work. 

    \item The individual star formation histories of galaxies in \flares\ are generally rising. However, the fraction with rising SFHs declines with redshift and stellar mass. Star formation histories are well described by a (two parameter) truncated normal or truncated log-normal distribution.
    
    \item There is a well defined stellar mass - metallicity relation in place at $z=10$ and beyond \citep[for $z>10$ see ][]{Wilkins2022a}. Galaxies at these redshifts are also predicted to be very $\alpha$-enhanced, with [Fe/H]$\approx 0.65$ at $z=5$. On the surface, these predictions are discrepant with recent metallicity measurements based on rest-frame observations of galaxies at $z = 2.5-5$ \citep{Cullen19, Calabro21}. However, the inferred iron mass fraction, which these observations are sensitive to, are much similar, particularly at high-masses where we have the most confidence in the models and observations.
    
    \item Within individual galaxies the metallicities of individual star particles are both inversely correlated with age and span a wide range, typically $1-1.5$ dex. 
    
    \item We find no evidence of an environmental dependence of the relationship between stellar mass, and metallicity, star formation, or age. However, this is not to say that galaxy formation is not sensitive to environment: the densest regions contain a (much) larger fraction of their stellar mass in massive metal enriched galaxies making these environments more metal enriched than lower density regions. 
    
\end{itemize}

At present our predictions are \emph{mostly} consistent with existing observational constraints, including the handful of observations to emerge so far from \webb. However, at least in part, this current good agreement reflects the large observational uncertainties due to small samples, limited $>2\mu$m observations (particularly spectroscopy), and observational modelling assumptions (e.g. the choice of stellar population synthesis model, or star formation history parameterisation). New observations from \webb, and in particular deep rest-frame optical spectroscopy, may soon challenge these predictions.

\section*{Acknowledgements}

We thank the \eagle\, team for their efforts in developing the \eagle\, simulation code.  This work used the DiRAC@Durham facility managed by the Institute for Computational Cosmology on behalf of the STFC DiRAC HPC Facility (\url{www.dirac.ac.uk}). The equipment was funded by BEIS capital funding via STFC capital grants ST/K00042X/1, ST/P002293/1, ST/R002371/1 and ST/S002502/1, Durham University and STFC operations grant ST/R000832/1. DiRAC is part of the National e-Infrastructure. CCL acknowledges support from the Royal Society under grant RGF/EA/181016. DI acknowledges support by the European Research Council via ERC Consolidator Grant KETJU (no. 818930). The Cosmic Dawn Center (DAWN) is funded by the Danish National Research Foundation under grant No. 140. EZ acknowledge funding from the Swedish National Space Agency. We also wish to acknowledge the following open source software packages used in the analysis: \textsc{Numpy} \citep{numpy:2020}, \textsc{Scipy} \citep{2020SciPy-NMeth}, and \textsc{Matplotlib} \citep{Hunter:2007}. This research made use of \textsc{Astropy} \url{http://www.astropy.org} a community-developed core Python package for Astronomy \citep{astropy:2013, astropy:2018}. Parts of the results in this work make use of the colormaps in the \textsc{CMasher} package \citep{CMasher}.
 
\section*{Data Availability}
 
Binned data (medians and ranges) are available in the \texttt{astropy} \href{https://github.com/astropy/astropy-APEs/blob/main/APE6.rst}{Enhanced Character Separated Values (\texttt{.ecsv}) table format} at \url{https://github.com/stephenmwilkins/flares_sfzh_data} and as part of the wider \href{}{First Light and Assembly of Galaxies} model predictions repository available at \url{https://github.com/stephenmwilkins/flags_data}. Data from the wider \flares\ project is available at \url{https://flaresimulations.github.io/data.html}. If you use data from this paper please also cite \citet{FLARES-I} and \citet{FLARES-II}.



\bibliographystyle{mnras}
\bibliography{flares_sfzh} 




\appendix

\section{Measuring stellar masses and star formation rates}

\subsection{Apertures}\label{sec:appendix.apertures}

\begin{figure*}
    \includegraphics[width=0.6\columnwidth]{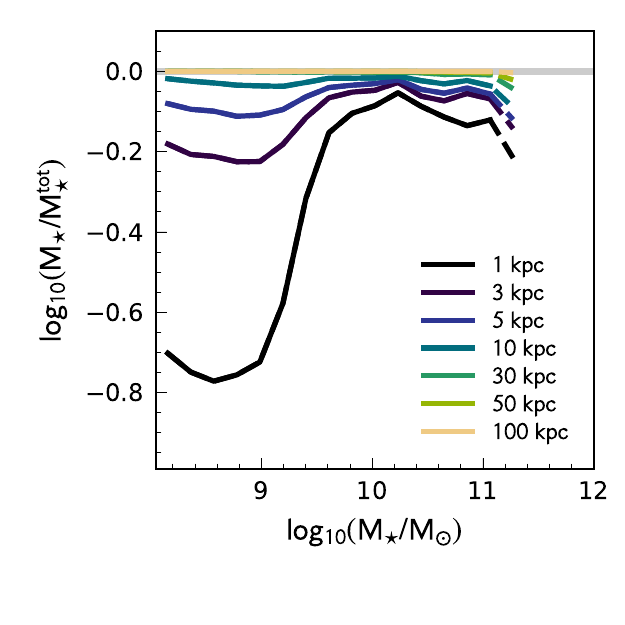}
    \includegraphics[width=0.6\columnwidth]{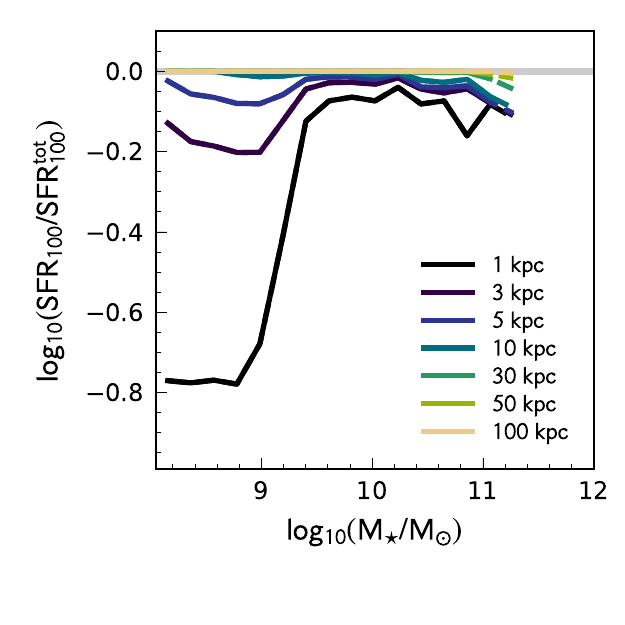}
	\includegraphics[width=0.6\columnwidth]{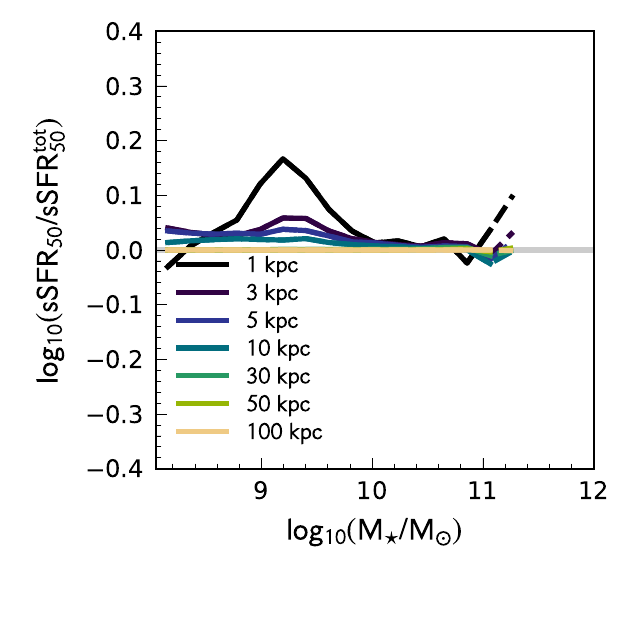}
	\caption{The sensitivity of stellar mass (left), star formation rate (centre), and specific star formation rate (right) to the choice of aperture used to define the galaxy. \label{fig:sim.aperture}}
\end{figure*}

Galaxy formation simulations identify objects as bound structures including halos and sub-halos. However, to better match observational methodologies, it is common to measure galaxy properties in a spherical aperture. In both the core \eagle\ and \flares\ analyses 30 kpc diameter apertures were used. In Fig. \ref{fig:sim.aperture} we briefly explore the consequences of this assumption on the stellar mass, star formation rate, and specific star formation rate of galaxies at $z=5$. While our chosen aperture captures virtually all stellar mass and star formation, smaller apertures tend to miss both in the lowest and highest mass galaxies. However, despite this sensitivity the measured specific star formation rate is mostly insensitive to the choice of aperture.

\subsection{Star formation measures}\label{sec:appendix.avgsfr}

\begin{figure}
	\includegraphics[width=\columnwidth]{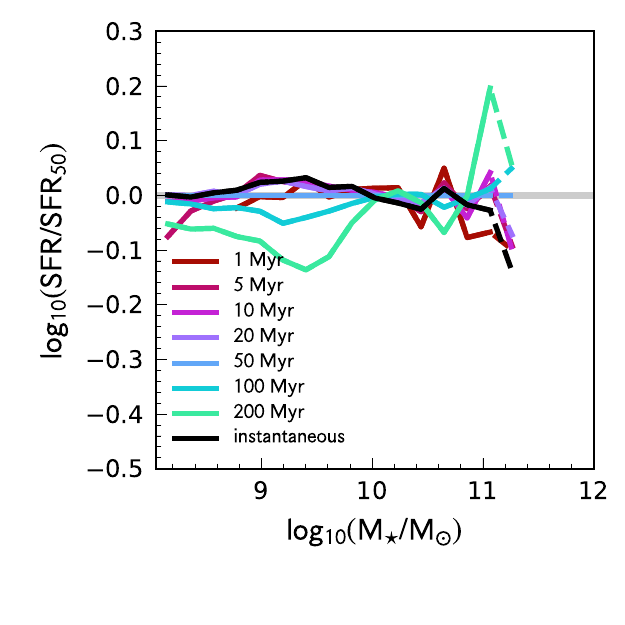}
	\caption{The impact of the star formation averaging timescale on star formation rates. \label{fig:sim.timescale}}
\end{figure}

Within hydrodynamical simulations like \flares\ there are two approaches for measuring star formation rates (SFRs): 1) from the properties of the gas particles themselves - the  \emph{instantaneous} SFR and 2) from the star particles by averaging over a particular timescale. Short timescales $<10$ Myr are limited by sampling effects: in low-SFR galaxies there may only be handful, or even no, star particles formed. While short-timescales are likely to match SF diagnostics based on reprocessed Lyman continuum emission (e.g. H$\alpha$) the same is \emph{possibly} not true for the FUV, and by extension far-IR, which have a contribution from stars up to several hundred million years old. In Figure \ref{fig:sim.timescale} we explore how the choice of averaging timescale affects the SFR. At high-mass ($M^{\star}>10^{10}\ {\rm M_{\odot}}$) the choice of timescale has little effect, the implication being that star formation histories in these galaxies are approximately constant. At lower masses however SFRs derived from longer timescales are typically smaller suggesting rising star formation histories. As expected instantaneous SFRs derived from gas particles closely match short-timescales. Unless otherwise stated, in this work we choose to average on a 50 Myr timescale. 

\bsp	
\label{lastpage}
\end{document}